\documentclass[aps,twocolumn,superscriptaddress,reprint]{revtex4-2}
\usepackage[colorinlistoftodos]{todonotes}
\usepackage{newtxtext,newtxmath,booktabs,siunitx} 
\usepackage{dcolumn}% Align table columns on decimal point
\usepackage{bm}% bold math
\usepackage{float}
\usepackage{subcaption}
\usepackage[]{graphicx}  % remove 'demo' option for your real document
\usepackage{booktabs}
\usepackage{tabularx}
\usepackage{amsmath}
\usepackage{xcolor}
\usepackage[colorlinks,citecolor=blue,urlcolor=blue,linkcolor=blue]{hyperref}

\DeclareSIUnit\eVperc{\eV\per\clight}
\DeclareSIUnit\clight{\text{\ensuremath{c}}}
\DeclareSIUnit\MeVA{A \mega\eV }

\newcommand{\mrm}{\mathrm}
\newcommand{\spirit}{S$\pi$RIT }

\newcolumntype{C}{>{\centering\arraybackslash}X}

\newcommand{\comment}[1]{}

\def\eqref#1{Eq.~(\ref{eq:#1})}

\makeatletter
\renewcommand\@makecaption[2]{%
  \par
  \vskip\abovecaptionskip
  \begingroup
   \small\rmfamily
    \begingroup
     \samepage
     \flushing
     \let\footnote\@footnotemark@gobble
     \@make@capt@title{#1}{#2}\par
    \endgroup
  \endgroup
  \vskip\belowcaptionskip
}
\makeatother

\graphicspath{ {images/} }

\begin{document}
%\begin{CJK*}{UTF8}{}
%\preprint{}

\title{Probing the Symmetry Energy with the Spectral Pion Ratio}

%\title{Pion Emission from Isospin Asymmetric Sn + Sn Collisions at 270 A MeV }
%\title{Charged Pion Emission in Neutron Rich Sn + Sn Collisions at 270 A MeV }
%\title{Charged Pion Emission in Neutron Rich Heavy Ion Collisions at 270 A MeV }
%\title{Pion Emission in Neutron Rich Heavy Ion Collisions at 270 A MeV }



\author{J.~Estee}
\altaffiliation{Corresponding author}
\email{esteejus@mit.edu}
\affiliation{National Superconducting Cyclotron Laboratory, Michigan State University, East Lansing, Michigan 48824, USA}
\affiliation{Department of Physics, Michigan State University, East Lansing, Michigan 48824, USA}

\author{W.G.~Lynch}%~(\CJKfamily{bsmi}{連致標})}
\altaffiliation{Corresponding author}
\email{lynch@nscl.msu.edu}
\affiliation{National Superconducting Cyclotron Laboratory, Michigan State University, East Lansing, Michigan 48824, USA}
\affiliation{Department of Physics, Michigan State University, East Lansing, Michigan 48824, USA}

\author{C.Y.~Tsang}%~(\CJKfamily{bsmi}{曾}}
\affiliation{National Superconducting Cyclotron Laboratory, Michigan State University, East Lansing, Michigan 48824, USA}
\affiliation{Department of Physics, Michigan State University, East Lansing, Michigan 48824, USA}

\author{J.~Barney}
\affiliation{National Superconducting Cyclotron Laboratory, Michigan State University, East Lansing, Michigan 48824, USA}
\affiliation{Department of Physics, Michigan State University, East Lansing, Michigan 48824, USA}

\author{G.~Jhang}%~(\CJKfamily{mj}{장진희})}
\affiliation{National Superconducting Cyclotron Laboratory, Michigan State University, East Lansing, Michigan 48824, USA}

\author{M.B.~Tsang}%~(\CJKfamily{bsmi}{曾敏兒})}
\altaffiliation{Corresponding author}
\email{tsang@nscl.msu.edu}
\affiliation{National Superconducting Cyclotron Laboratory, Michigan State University, East Lansing, Michigan 48824, USA}
\affiliation{Department of Physics, Michigan State University, East Lansing, Michigan 48824, USA}

\author{R.~Wang}%~(\CJKfamily{gbsn}{王})}
\affiliation{National Superconducting Cyclotron Laboratory, Michigan State University, East Lansing, Michigan 48824, USA}

\author{M.~Kaneko}%~(\CJKfamily{min}{金子雅紀})}
\affiliation{RIKEN Nishina Center, Hirosawa 2-1, Wako, Saitama 351-0198, Japan}
\affiliation{Department of Physics, Kyoto University, Kita-shirakawa, Kyoto 606-8502, Japan}

\author{J.W.~Lee}%~(\CJKfamily{mj}{이정우})}
\affiliation{Department of Physics, Korea University, Seoul 02841, Republic of Korea}

\author{T.~Isobe}%~(\CJKfamily{min}{磯部忠昭})}
\altaffiliation{Corresponding author}
\email{isobe@riken.jp}
\affiliation{RIKEN Nishina Center, Hirosawa 2-1, Wako, Saitama 351-0198, Japan}

\author{M.~Kurata-Nishimura}%~(\CJKfamily{min}{倉田-西村美月})}
\affiliation{RIKEN Nishina Center, Hirosawa 2-1, Wako, Saitama 351-0198, Japan}

\author{T.~Murakami}%~(\CJKfamily{min}{村上哲也})}
\altaffiliation{Corresponding author}
\email{murakami.tetsuya.3e@kyoto-u.ac.jp}
\affiliation{RIKEN Nishina Center, Hirosawa 2-1, Wako, Saitama 351-0198, Japan}
\affiliation{Department of Physics, Kyoto University, Kita-shirakawa, Kyoto 606-8502, Japan}

\author{D.S.~Ahn}%~(\CJKfamily{mj}{안득순})}
\affiliation{RIKEN Nishina Center, Hirosawa 2-1, Wako, Saitama 351-0198, Japan}

\author{L.~Atar}
\affiliation{Institut f\"ur Kernphysik, Technische Universit\"at Darmstadt, D-64289 Darmstadt, Germany}
\affiliation{GSI Helmholtzzentrum f\"ur Schwerionenforschung, Planckstrasse 1, 64291 Darmstadt, Germany}

\author{T.~Aumann}
\affiliation{Institut f\"ur Kernphysik, Technische Universit\"at Darmstadt, D-64289 Darmstadt, Germany}
\affiliation{GSI Helmholtzzentrum f\"ur Schwerionenforschung, Planckstrasse 1, 64291 Darmstadt, Germany}

\author{H.~Baba}%~(\CJKfamily{min}{馬場秀忠})}
\affiliation{RIKEN Nishina Center, Hirosawa 2-1, Wako, Saitama 351-0198, Japan}

\author{K.~Boretzky}
\affiliation{GSI Helmholtzzentrum f\"ur Schwerionenforschung, Planckstrasse 1, 64291 Darmstadt, Germany}

\author{J.~Brzychczyk}
\affiliation{Faculty of Physics, Astronomy and Applied Computer Science, Jagiellonian University, Krak\'ow, Poland}

\author{G.~Cerizza}
\affiliation{National Superconducting Cyclotron Laboratory, Michigan State University, East Lansing, Michigan 48824, USA}

\author{N.~Chiga}
\affiliation{RIKEN Nishina Center, Hirosawa 2-1, Wako, Saitama 351-0198, Japan}

\author{N.~Fukuda}
\affiliation{RIKEN Nishina Center, Hirosawa 2-1, Wako, Saitama 351-0198, Japan}

\author{I.~Gasparic}
\affiliation{Division of Experimental Physics, Rudjer Boskovic Institute, Zagreb, Croatia}
\affiliation{RIKEN Nishina Center, Hirosawa 2-1, Wako, Saitama 351-0198, Japan}
\affiliation{Institut f\"ur Kernphysik, Technische Universit\"at Darmstadt, D-64289 Darmstadt, Germany}

\author{B.~Hong}%~(\CJKfamily{mj}{홍병식})}
\affiliation{Department of Physics, Korea University, Seoul 02841, Republic of Korea}

\author{A.~Horvat}
\affiliation{Institut f\"ur Kernphysik, Technische Universit\"at Darmstadt, D-64289 Darmstadt, Germany}
\affiliation{GSI Helmholtzzentrum f\"ur Schwerionenforschung, Planckstrasse 1, 64291 Darmstadt, Germany}

\author{K.~Ieki}
\affiliation{Department of Physics, Rikkyo University, Nishi-Ikebukuro 3-34-1, Tokyo 171-8501, Japan}

\author{N.~Inabe}
\affiliation{RIKEN Nishina Center, Hirosawa 2-1, Wako, Saitama 351-0198, Japan}

\author{Y.J.~Kim}%~(\CJKfamily{mj}{김영진})
\affiliation{Rare Isotope Science Project, Institute for Basic Science, Daejeon 34047, Republic of Korea}

\author{T.~Kobayashi}
\affiliation{Department of Physics, Tohoku University, Sendai 980-8578, Japan}

\author{Y.~Kondo}
\affiliation{Department of Physics, Tokyo Institute of Technology, Tokyo 152-8551, Japan}

\author{P.~Lasko}
\affiliation{Institute of Nuclear Physics PAN, ul. Radzikowskiego 152, 31-342 Krak\'ow, Poland}

\author{H.~S.~Lee}%~(\CJKfamily{mj}{이효상})}
\affiliation{Rare Isotope Science Project, Institute for Basic Science, Daejeon 34047, Republic of Korea}

\author{Y.~Leifels}
\affiliation{GSI Helmholtzzentrum f\"ur Schwerionenforschung, Planckstrasse 1, 64291 Darmstadt, Germany}

\author{J.~\L{}ukasik}
\affiliation{Institute of Nuclear Physics PAN, ul. Radzikowskiego 152, 31-342 Krak\'ow, Poland}

\author{J.~Manfredi}
\affiliation{National Superconducting Cyclotron Laboratory, Michigan State University, East Lansing, Michigan 48824, USA}
\affiliation{Department of Physics, Michigan State University, East Lansing, Michigan 48824, USA}

\author{A.~B.~McIntosh}
\affiliation{Cyclotron Institute, Texas A\&M University, College Station, Texas 77843, USA}

\author{P.~Morfouace}
\affiliation{National Superconducting Cyclotron Laboratory, Michigan State University, East Lansing, Michigan 48824, USA}

\author{T.~Nakamura}
\affiliation{Department of Physics, Tokyo Institute of Technology, Tokyo 152-8551, Japan}

\author{N.~Nakatsuka}%~(\CJKfamily{min}{中塚徳継})}
\affiliation{RIKEN Nishina Center, Hirosawa 2-1, Wako, Saitama 351-0198, Japan}
\affiliation{Department of Physics, Kyoto University, Kita-shirakawa, Kyoto 606-8502, Japan}

\author{S.~Nishimura}%~(\CJKfamily{min}{西村俊二})}
\affiliation{RIKEN Nishina Center, Hirosawa 2-1, Wako, Saitama 351-0198, Japan}

\author{H.~Otsu}
\affiliation{RIKEN Nishina Center, Hirosawa 2-1, Wako, Saitama 351-0198, Japan}

\author{P.~Paw\l{}owski}
\affiliation{Institute of Nuclear Physics PAN, ul. Radzikowskiego 152, 31-342 Krak\'ow, Poland}

\author{K.~Pelczar}
\affiliation{Faculty of Physics, Astronomy and Applied Computer Science, Jagiellonian University, Krak\'ow, Poland}

\author{D.~Rossi}
\affiliation{Institut f\"ur Kernphysik, Technische Universit\"at Darmstadt, D-64289 Darmstadt, Germany}

\author{H.~Sakurai}%~(\CJKfamily{min}{櫻井博義})}
\affiliation{RIKEN Nishina Center, Hirosawa 2-1, Wako, Saitama 351-0198, Japan}

\author{C.~Santamaria}
\affiliation{National Superconducting Cyclotron Laboratory, Michigan State University, East Lansing, Michigan 48824, USA}

\author{H.~Sato}
\affiliation{RIKEN Nishina Center, Hirosawa 2-1, Wako, Saitama 351-0198, Japan}

\author{H.~Scheit}
\affiliation{Institut f\"ur Kernphysik, Technische Universit\"at Darmstadt, D-64289 Darmstadt, Germany}

\author{R.~Shane}
\affiliation{National Superconducting Cyclotron Laboratory, Michigan State University, East Lansing, Michigan 48824, USA}

\author{Y.~Shimizu}
\affiliation{RIKEN Nishina Center, Hirosawa 2-1, Wako, Saitama 351-0198, Japan}

\author{H.~Simon}
\affiliation{GSI Helmholtzzentrum f\"ur Schwerionenforschung, Planckstrasse 1, 64291 Darmstadt, Germany}

\author{A.~Snoch}
\affiliation{Nikhef National Institute for Subatomic Physics, Amsterdam, Netherlands}

\author{A.~Sochocka}
\affiliation{Faculty of Physics, Astronomy and Applied Computer Science, Jagiellonian University, Krak\'ow, Poland}

%\author{Z.~Sosin\footnote{Deceased}}
%\affiliation{Faculty of Physics, Astronomy and Applied Computer Science, Jagiellonian University, Krak\'ow, Poland}

\author{T.~Sumikama}
\affiliation{RIKEN Nishina Center, Hirosawa 2-1, Wako, Saitama 351-0198, Japan}

\author{H.~Suzuki}
\affiliation{RIKEN Nishina Center, Hirosawa 2-1, Wako, Saitama 351-0198, Japan}

\author{D.~Suzuki}
\affiliation{RIKEN Nishina Center, Hirosawa 2-1, Wako, Saitama 351-0198, Japan}

\author{H.~Takeda}
\affiliation{RIKEN Nishina Center, Hirosawa 2-1, Wako, Saitama 351-0198, Japan}

\author{S.~Tangwancharoen}
\affiliation{National Superconducting Cyclotron Laboratory, Michigan State University, East Lansing, Michigan 48824, USA}

\author{H.~Toernqvist}
\affiliation{Institut f\"ur Kernphysik, Technische Universit\"at Darmstadt, D-64289 Darmstadt, Germany}
\affiliation{GSI Helmholtzzentrum f\"ur Schwerionenforschung, Planckstrasse 1, 64291 Darmstadt, Germany}

\author{Y.~Togano}
\affiliation{Department of Physics, Rikkyo University, Nishi-Ikebukuro 3-34-1, Tokyo 171-8501, Japan}

\author{Z.~G.~Xiao}
\affiliation{Department of Physics, Tsinghua University, Beijing 100084, PR China}

\author{S.~J.~Yennello}
\affiliation{Cyclotron Institute, Texas A\&M University, College Station, Texas 77843, USA}
\affiliation{Department of Chemistry, Texas A\&M University, College Station, Texas 77843, USA}

\author{Y.~Zhang}
\affiliation{Department of Physics, Tsinghua University, Beijing 100084, PR China}
\collaboration{\spirit collaboration}

%\author {and}

\author{M.D.~Cozma}
\affiliation{National Superconducting Cyclotron Laboratory, Michigan State University, East Lansing, Michigan 48824, USA}
\affiliation{IFIN-HH, Reactorului 30, 077125 M\v{a}gurele-Bucharest, Romania}

%\author{P.~Danielewicz}
%%\author{D.~Oliinychenko}
%\author{H.~Elfnerg}
%\author{N.~Ikeno}
%\author{C.M.~Ko}
%\author{J. Moh}
%\author{A.~Ono}
%\author{J.~Su}
%\author{Yong Jia Wang}
%\author{H.~Wolter}
%\author{Zhen Zhang}
%\author{Ying-Xun Zhang}
%\collaboration{TMEP collaboration}


\date{\today}

\begin{abstract}
 Many neutron star (NS) properties, such as the proton fraction within a NS, reflect the symmetry energy contributions to the Equation of State that dominate when neutron and proton densities differ strongly. To constrain these contributions at supra-saturation densities, we measure the spectra of charged pions produced by colliding rare isotope tin (Sn) beams with isotopically enriched Sn targets. Using ratios of the charged pion spectra measured at high transverse momenta, we deduce the slope of the symmetry energy to be $42 < L < \SI{117}{MeV}$. This value is slightly lower but consistent with the $L$ values deduced from a recent measurement of the neutron skin thickness of $^{208}$Pb.
 %and the symmetry pressure to be $17\pm\SI{15e33}{dyne\per\centi\meter\squared}$ at a density of
%$\SI{3.6e14}{\gram\per\centi\metre\cubed}$.

\end{abstract}

%\keywords{}

\maketitle
%\end{CJK*}

Recent gravitational wave measurements of the neutron star merger event GW170817 provide information about the deformability of neutron stars (NS)~\cite{Abb17, Abb18}. Analyses of the gravitational wave signal reveal that this deformability mainly reflects the nuclear Equation of State (EoS) at densities of about twice the saturation density of nuclear matter, $\rho_0 \approx \SI{2.4e14}{\gram\per\centi\metre\cubed}$ or $\SI{0.16}{\per\femto\metre\cubed}$. While the GW170817 observations provide key insights into NS and their mergers, they do not reveal how the NS EoS depends on the abundances of its constituent neutrons, protons, $\Delta$ resonances, and pions~\cite{Tsa19b,Tsa19,Lim18,Tews18,Lat01,Dra14,Duc11,Lat16,Jli19,For20}.  To understand what is the prevailing form of matter in the NS outer core, such microscopic information is essential.

  Microscopic information about the EoS has only been extracted from laboratory experiments. Measurements of nucleus-nucleus collisions have constrained the EoS for symmetric matter comprised of equal proton, $\rho_p$, and neutron, $\rho_n$, densities for total densities  $\rho = \rho_n+\rho_p$ of $\rho_0 \leq \rho \leq 4.5\rho_0$~\cite{Dan02,Lyn09,LeF16}.  The main challenge at  $\rho > \rho_0$ is to understand the symmetry energy, which describes how the EoS depends upon isovector potentials that have the opposite sign for neutrons as for protons and depend linearly on the difference between neutron and proton densities $(\rho_n - \rho_p)$, or equivalently on the isospin asymmetry  $\delta  =(\rho_n - \rho_p)/\rho$~\cite{Hor14,Tsa19b,Tsa10,Duc11,Li08,Lat01}. 

The symmetry energy has been constrained at sub-saturation densities using a variety of nuclear structure and reaction observables~\cite{Hor14,Tsa10,lynch2018nuclear}. To probe higher densities, one must study central collisions between two complex nuclei. At incident energies of about \SI{300}{\MeVA} and above, nuclear matter can be compressed to densities approaching 2$\rho_0 $~\cite{Ike16}. The isovector mean field potentials cause the flow of neutrons emitted from this dense region to differ from the flow of protons; this difference provides an observable that can constrain the symmetry energy~\cite{Coz18,Li08}.

In these dense regions, nucleon-nucleon inelastic collisions produce $\Delta$ baryons that decay to nucleons by emitting pions. From the $\Delta$ production and decay cross sections, one expects the ratio  $\mrm{M}(\pi^-)/\mrm{M}(\pi^+)$ of the multiplicity (M) of negatively and positively charged pions per collision to be proportional to $(\rho_n / \rho_p)^2$~\cite{bao_pi,Li08}. Because the ratio $(\rho_n / \rho_p)$ strongly reflects the isovector mean field potentials within this dense region, both the total pion multiplicity yield ratio $\mrm{M}(\pi^-)/\mrm{M}(\pi^+)$~\cite{Li08,Bal05} and the dependence of the pion ratio on pion momentum~\cite{Bal05,Hon14,Tsa17}, reflect the density dependence of the symmetry energy. Existing studies of $\mrm{M}(\pi^-)/\mrm{M}(\pi^+)$~\cite{Xia09,Coz16} with stable nuclear beams have not provided a consistent constraint on the symmetry energy at supra-saturation densities, $\rho>\rho_0$. This may result from different assumptions for the $\Delta$ and pion potentials that cause the calculated low energy pion spectra, the $\mrm{M}(\pi^-)/\mrm{M}(\pi^+)$ ratios and the symmetry energy constraints to differ~\cite{Cozma:2020bre}.

Powerful new radioactive isotope facilities are being built to investigate how nuclei and the nuclear EoS depend on $\delta=(\rho_n - \rho_p)/\rho$~\cite{Blu13,Li08,Hor14}.  Here we present results from the first experiment at these new facilities to probe the symmetry energy at high density with radioactive beams. In this experiment, beams of ${}^{132}$Sn, ${}^{124}$Sn, ${}^{112}$Sn, and ${}^{108}$Sn projectiles at \SI{270}{\MeVA} incident energy bombarded isotopically enriched (>95\%) ${}^{124}$Sn and ${}^{112}$Sn  targets of 608 and \SI{561}{\milli\gram\per\centi\metre\squared} areal density at the Radioactive Ion Beam Factory (RIBF) in Japan. Light charged particles, including $\pi^-$ and  $\pi^+$ mesons, were detected in a new device, the  S$\pi$RIT Time Projection Chamber (TPC), placed inside the SAMURAI spectrometer~\cite{Ots16}. 
 
 Previous publications describe the design and performance of the S$\pi$RIT TPC~\cite{Sha15,Tan17,Bar20},  its trigger systems~\cite{Las17},  electronics readout system~\cite{Iso18}, and analysis software~\cite{Lee20, Github}. To measure minimum ionizing pions as well as isotopically resolved H, He, and Li isotopes, we expanded the typical dynamic range of the TPC electronics by a factor of 5~\cite{Est19}. To seamlessly measure pions over the essential range of scattering angles, we placed the target at the entrance of the TPC and corrected for space charge effects from the beam traversing the TPC~\cite{Tsa20}.
 
Charged particles were identified by their electronic stopping powers $\mrm{dE/dx}$ and magnetic rigidities~\cite{Lee20}.  To optimize momentum resolution, pion data are measured at azimuthal angles  $-\ang{40} < \phi < \ang{25} \cup \ang{160} < \phi < \ang{210} $, where the pion momenta are mainly perpendicular to the magnetic field.  Clean pion identification was achieved. We utilize $\phi$ independence and interpolate the pion spectra to other azimuthal angles where the momentum resolution would be inferior. We fit the $\mrm{dE/dx}$ distributions for each momentum bin to determine the particle yield and the background contribution. 

We focus on the most central collisions with the highest charged particle multiplicities corresponding to impact parameters of $b<\SI{3}{fm}$~\cite{Bar20}. Electrons and positrons from the Dalitz decay of $\pi^0$ are the largest contributions to the pion background and have been subtracted as detailed in Ref.~\cite{Bar19}. These background contributions are insignificant.

The TPC pion acceptance in the current experiment allows energy of pions to be accurately measured down to \SI{0}{\mega \electronvolt} in the center-of-mass of the total system (CM). We focus on pions measured to polar angles of $\theta_\text{CM}~<~\ang{90}$ with respect to the beam where pion acceptance is complete. This angular cut is also applied to the theoretical calculations discussed later. We calculate the efficiency by embedding Monte-Carlo pion tracks into real events and determining the fraction of these tracks that are accurately reconstructed. We used a calibration beam composed of hydrogen isotopes at well known momenta to check  the momentum determination of the TPC. The momentum values obtained by using the TPC design geometry and SAMURAI dipole magnetic field agreed to within 1\% of the known values~\cite{Est20}. The estimated systematic uncertainties are  4\% for the individual pion spectra and  2\% for the single and double ratios of charged pion spectra. These uncertainties are incorporated into the discussion below. 
 
 The total $\pi^-$ and $\pi^+$ multiplicities and their ratios for central ($b\approx\SI{3}{\femto\metre}$) ${}^{132}$Sn + ${}^{124}$Sn, ${}^{112}$Sn + ${}^{124}$Sn, ${}^{108}$Sn + ${}^{112}$Sn collisions are published in Ref.~\cite{Jhang:JHA20}. Comparisons of the total pion ratios predicted by seven different theoretical calculations exhibit differences among them that exceed their sensitivity to the symmetry energy. Different assumptions regarding the mean field potentials for $\Delta$ baryons and pions can strongly influence the production of low energy pions and thus the total charged pion multiplicities and their ratios~\cite{Cozma:2020bre}. To reduce this sensitivity, we focus on pion spectra at higher momenta where sensitivity to the isospin dependence of the nucleonic mean fields dominates~\cite{Cozma:2020bre}. Using the pion  spectral ratios at high transverse momenta, we obtain a correlated constraint at supra-saturation densities on  the symmetry energy and the momentum dependence of the isovector nucleonic mean field potentials.
 
 For our investigations, we use the dcQMD semi-classical Quantum Molecular Dynamics (QMD) model of Ref.~\cite{Cozma:2020bre}. This model has provided constraints on the symmetry energy from neutron and proton elliptic flow measurements~\cite{Coz18} and from pion production~\cite{Coz16,Coz17}. It also provides reasonable predictions of the pion multiplicities and ratios for the current experiment~\cite{Jhang:JHA20}. A unique aspect of the dcQMD model is the implementation of the conservation of total energy for the system, which is not simply satisfied at the two-body level due to the momentum and isospin asymmetry dependence of interactions. This involves modifying the collision term to allow for energy transfer between scattering particles and the rest of the system, leading to shifts of particle production thresholds~\cite{Ferini:2006je,Song:2015hua,Coz16}. With this correction, consistent constraints for the symmetry energy density dependence were obtained from pion production and elliptic flow~\cite{Coz16}. Further details of this model can be found in Refs.~\cite{Coz16,Coz17,Coz18,Cozma:2020bre}.

At beam energies of \SI{270}{\MeVA}, high energy pions are primarily produced by exciting $\Delta$ (1232) baryons  via two-nucleon $N+N\rightarrow N+\Delta$ inelastic scattering processes. These $\Delta$'s may scatter elastically or inelastically via $N+\Delta\rightarrow N+\Delta '$ or decay via $\Delta\rightarrow N +\pi$ producing pions. Pions, in turn, can be absorbed via $\pi+N\rightarrow \Delta$.  Details of the $\Delta$ resonance production parameterization and its modification in nuclear medium can be found in Refs.~\cite{Huber:1994ee,Larionov:2003av}. The present calculations require realistic binding energies per nucleon, charged radii and neutron skins for projectile and target nuclei and a good quantitative description of the experimental stopping, directed flow and elliptic flow observables~\cite{Reisdorf:2010aa,FOPI:2011aa}. These prior analyses are consistent with the isoscalar effective mass $m^*/m=0.7$, compressibility modulus $K_0=\SI{245}{\mega\electronvolt}$~\cite{Cozma:2020bre} and in-medium elastic nucleon-nucleon cross-sections~\cite{Li:2011zzp} used here.

The Gaussian wave functions for nucleons and pions in dcQMD have widths that reflect the experimental ratio of pion-to-proton charge radii~\cite{Coz17}. Pions move under the  influence of the Coulomb interaction and S and P wave pion optical potentials calculated with the ``Batty-1''  parameters of Ref.~\cite{Coz17,Batty:1978aa}. We find that the usual Ansatz of setting the $\Delta$ potential in nuclear matter equal to that of nucleons leads to incorrect $\pi^-$ and $\pi^+$ production thresholds and total multiplicities \cite{Cozma:2020bre,Jhang:JHA20}. Therefore, we adjust the potential depths at saturation density and effective masses in the iso-scalar and iso-vector channels~\cite{Cozma:2020bre} of the  $\Delta$ to reproduce experimental total pion multiplicities and mean kinetic energies.  

In dcQMD, the nuclear EoS is defined in terms of the energy per nucleon and is given by~\cite{Das:2002fr}: 

\begin{multline}
\label{eos}
\frac{E}{N} (\rho,\delta) = \text{KE}(\rho,\delta)+ A_u \frac{\rho(1-\delta^2)}{4\rho_0} + A_l\frac{\rho(1+\delta^2)}{4\rho_0} \\
 + \frac{B}{\sigma+1} \frac{\rho^{\sigma}}{\rho_0^\sigma}(1-x\delta^2) +
\frac{D}{3}\frac{\rho^2}{\rho_0^2}(1-y\delta^2) \\
 +\frac{1}{\rho\rho_0}\sum_{\tau,\tau'} C_{\tau \tau'}\int\int d^3\vec{p} d^3\vec{p}' 
\frac{f_\tau(\vec{r},\vec{p}) f_{\tau'}(\vec{r},\vec{p}')}{1+(\vec{p}-\vec{p}')^2/\Lambda^2}.
\end{multline}

Here, $\text{KE}(\rho,\delta)$ is the kinetic energy density, followed by four local potential energy terms that depend on density $\rho$ and asymmetry $\delta$. The final non-local term models Pauli exchange terms and the finite range of nucleon-nucleon interactions.   Parameter $D$ controls the compressibility $K_0=\SI{245}{MeV}$ and skewness $Q_0=\SI{-350}{MeV}$ of symmetric matter, and $x$ and $y$ controls slope $L$ and curvature $K_\text{sym}$ parameters of the symmetry energy $S(\rho)$. We correlate $L$ and $K_\text{sym}$ via $K_\text{sym}=-488+6.728 \times L$ (MeV) and also set $S(\rho = \SI{0.1}{\per\femto\metre\cubed} ) = \SI{25.5}{\mega\electronvolt}$, consistent with nuclear mass and radius measurements~\cite{Brown:2013mga,Zhang:2013wna,Trippa:2008g}. 

 \begin{figure}[ht!]
  \includegraphics[width=1.05\linewidth]{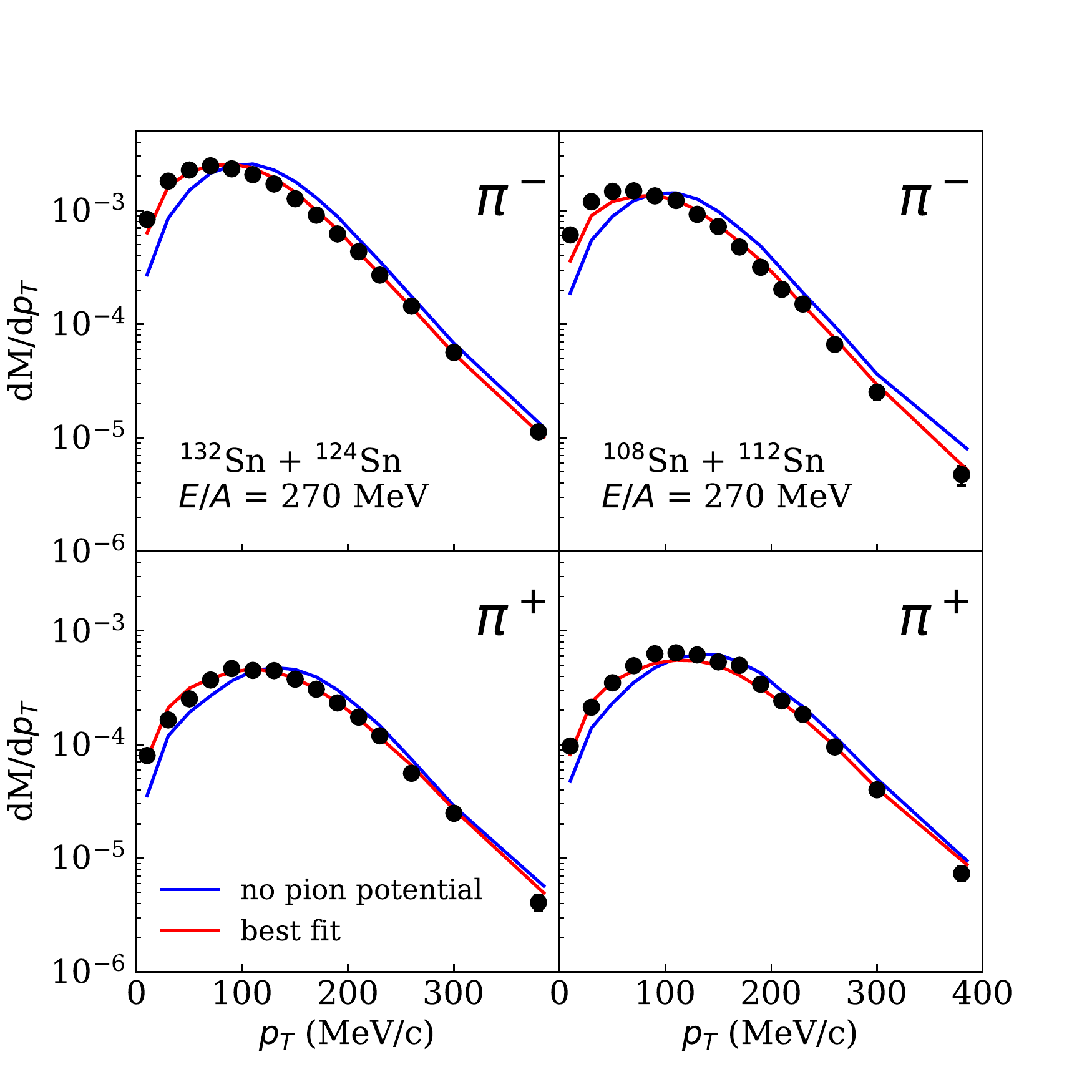}	
  \caption{Measured and calculated pion spectra. The red lines are the calculated pion spectra after adjusting the $\Delta$ potential to reproduce the pion multiplicities. The blue lines differ from the red lines in that the the pion optical potential has been removed. The nucleon potentials in these simulations correspond to $L=\SI{80}{MeV}$ and $\Delta m^*_{np}=0$.} 
  \label{fig:pionmultspectra}
\end{figure}

  The left and right panels of Fig.~\ref{fig:pionmultspectra} show the CM  transverse momentum spectra $\mrm{dM/d}p_{T}$ at $\theta_\text{CM}~<~\ang{90}$ for the very neutron rich $\mrm{ {}^{132}Sn + {}^{124}Sn }$ and the nearly symmetric $\mrm{ {}^{108}Sn + {}^{112}Sn }$ systems, respectively. The difference in the $p_{T}$ values for the maxima of the $\pi^-$ and $\pi^+$ spectra reflects the influence of the Coulomb interaction. The calculations with $L=\SI{80}{MeV}$, shown in the figure have been fitted to the total multiplicities by optimizing the $\Delta$ potentials and effective masses. Here, the scaled difference between neutron and proton effective masses, $ \Delta m^*_{np} /\delta = [m^*_{n}-m^*_{p}]/(m\delta)$ is set to zero. The red curves show the resulting calculations including pion optical potentials while the blue curves show calculations where pion potentials are removed. Simulations without the pion optical potential result in a significant underprediction of the pion spectra at low $p_{T}$ and that extends over a larger range of momenta in the case of the $\pi^+$ spectra in both reactions. However, the shapes of the spectra at higher transverse momentum $p_{T}>$\SI{200}{\mega\eV\per\clight} are largely unchanged by the choice of $\Delta$ and pion optical potentials, and remain sensitive to the nucleonic mean field potentials and to the symmetry energy~\cite{Cozma:2020bre}. Such sensitivities to the details in pion and $\Delta$ potentials for the low energy pions could account for the differences in transport code predictions for the total pion yields reported in Ref.~\cite{Jhang:JHA20}.
  
\begin{figure}
  \includegraphics[trim=0 1.5cm 0 1.5cm, clip, width=0.9\linewidth]{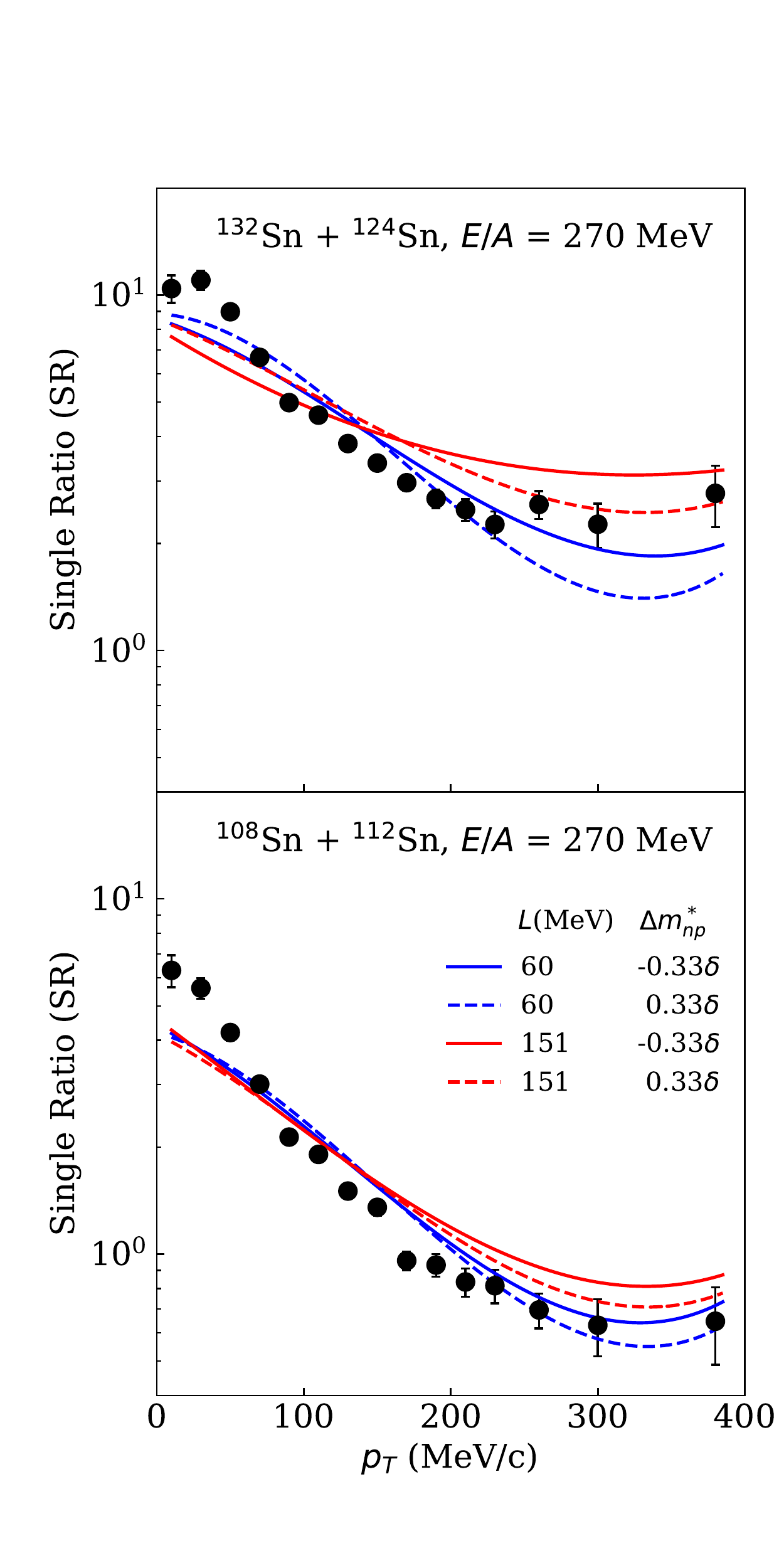}	
  \caption{Single pion spectral ratios for ${}^{132}$Sn$+{}^{124}$Sn (top panel) and ${}^{108}$Sn$+{}^{112}4$Sn (bottom panel) reactions. The curves are dcQMD predictions from different $L$ and $\Delta m^*_{np}$ values listed in the top panel.} 
  \label{fig:pmrspectra}
\end{figure}
  
Next, we focus on the isovector mean field potentials that contribute to the symmetry energy and are  opposite in sign for neutrons vs. protons and  $\pi^-$ vs. $\pi^+$.  We highlight these isovector potentials by constructing the single ratio $\mrm{ SR(\pi^-/\pi^+)=[dM(\pi^-)/d}p_{T}]/\mrm{[dM(\pi^+)/d}p_{T}] $. 
In Fig.~\ref{fig:pmrspectra}, $\mrm{ SR(\pi^-/\pi^+)_{132+124} }$ for the neutron rich $\mrm{ {}^{132}Sn+{}^{124}Sn }$ system is shown in the top panel and  $\mrm{ SR(\pi^-/\pi^+)_{108+112} }$ for the nearly symmetric $\mrm{ {}^{108}Sn+{}^{112}Sn}$ systems in the bottom panel. The steep rise in the single ratios at low $p_{T}$ originates from the opposite Coulomb forces experienced by $\pi^-$ and $\pi^+$.

 We construct the pion single spectral ratios using dcQMD with 12 sets of calculations with values for $L$ of (15, 60, 106 and \SI{151}{MeV}) and $ \Delta m^*_{np} /\delta$ of (-0.33, 0 and 0.33). For clarity, we show only four calculations with ($L$, $\Delta m^*_{np}/\delta$) = (60, -0.33), (60, 0.33), (151, -0.33) and (151, 0.33) represented by blue solid, blue dashed, red solid and red dashed curves respectively. All calculations under-predict the data at $p_{T} < \SI{50}{\mega\electronvolt\per\clight}$ and over-predict the data at $p_{T}\approx \SI{150}{\mega\electronvolt\per\clight}$ for both systems. As expected, the neutron rich system of $\mrm{{}^{132} Sn + {}^{124} Sn}$ displays much more sensitivity at high $p_{T}$ to the slope of the symmetry energy, $L$, than does the nearly symmetric $\mrm{{}^{108} Sn + {}^{112} Sn}$ system. 
The disagreement with data  observed at low $p_{T}$ for both systems suggests some   inaccuracy in the theory that does not depend strongly on the asymmetry $\delta$. Such effects could originate from inaccuracies in the treatment of Coulomb interactions or of the pion optical potentials above saturation density, for example. Non-resonant pion emission or absorption, neglected in the current calculations, could also contribute to the incorrect shape  of the single spectral ratios at low $p_{T}$ in Fig.~\ref{fig:pmrspectra} and its influence should be investigated. These effects should be much less important above \SI{200}{\mega\electronvolt\per\clight} where the trends of the data and the calculations become more comparable.

\begin{figure}
  \includegraphics[width=0.85\linewidth]{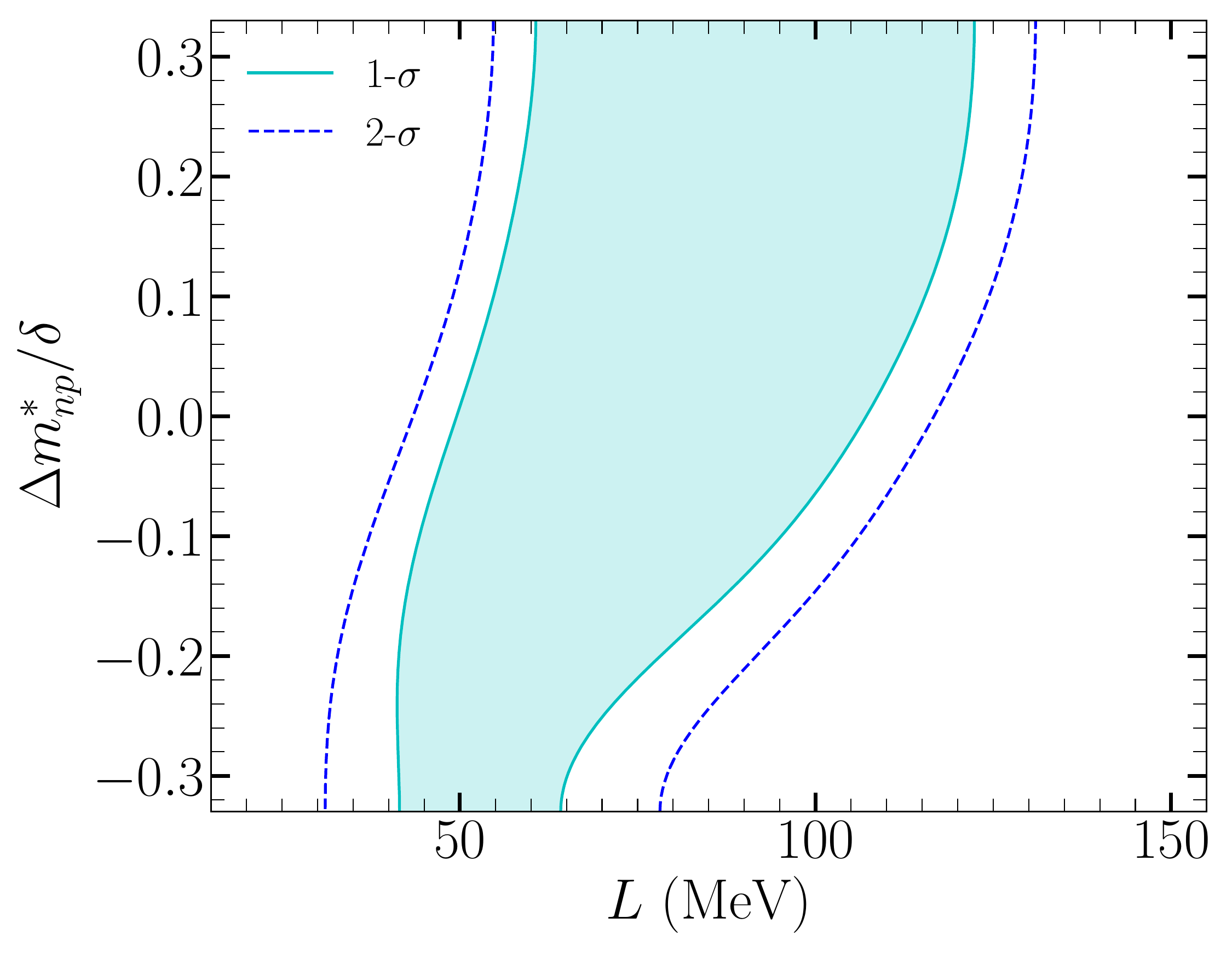}	
  \caption{Correlation contours between $L$ and $\Delta m^*_{np}/\delta$ extracted from the single pion spectral ratio of the neutron-rich ${}^{132}$Sn+${}^{124}$Sn and near symmetric  ${}^{108}$Sn+${}^{112}$Sn reactions. The green shaded region lies within the 68\% confidence level for data with $p_{T} > \SI{200}{MeV/c}$. The dotted blue lines denote contours corresponding to the 95\% confidence level. } 
  \label{fig:fitlsymnpemd}
\end{figure}

  Interpolating the dcQMD calculations, we fit the single ratios at $p_{T} > \SI{200}{\mega\electronvolt\per\clight}$ and extract correlated constraints on $L$ and $ \Delta m^*_{np} /\delta $ shown in Fig.~\ref{fig:fitlsymnpemd}. The correlated nature of this constraint means that larger values for  $ \Delta m^*_{np} $ would imply larger values for $L$. Absent any constraint on $ \Delta m^*_{np} $, the best fit value is $L=79.9 \pm \SI{37.6}{MeV}$ with  $S_0=35.3\pm\SI{2.8}{MeV}$, with statistical uncertainty making the largest contribution to the total uncertainty. 
  This value is consistent with constraints extracted from proton and neutron elliptic flows in Ref.~\cite{Coz18} using the same transport model. 

\begin{figure}
  \includegraphics[width=0.9\linewidth]{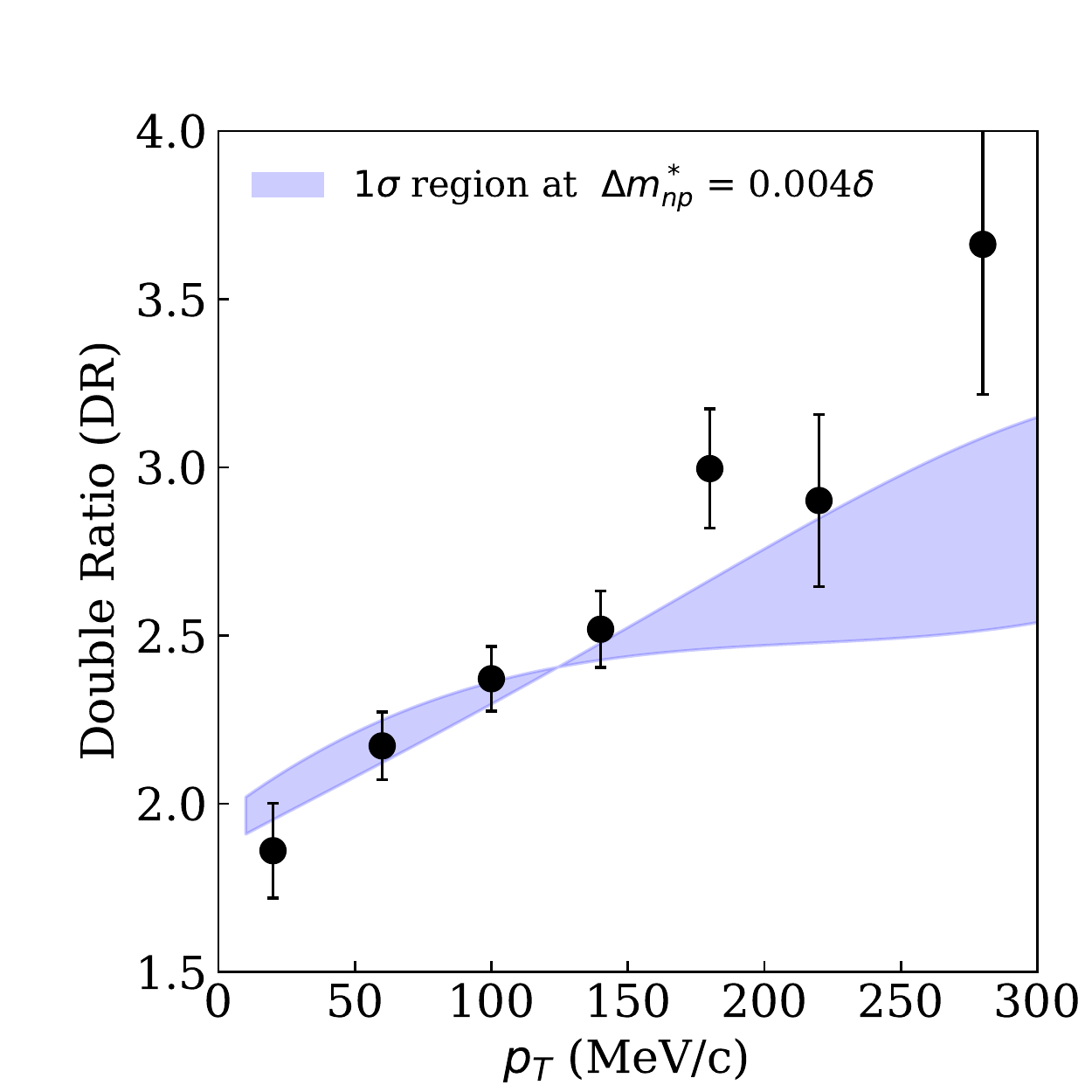}	
  \caption{Transverse momentum spectra of the double pion ratio. The shaded region covers dcQMD predictions within 1$\sigma$ of the most probable values of $L$ and $\Delta m^*_{np}$ values.} 
  \label{fig:drspectra}
\end{figure}

 Since both reactions have the same total charge, approximately the same isoscalar fields and differ principally by their asymmetry $\delta$, the double ratio, $\mrm{DR(\pi^-/\pi^+)=SR(\pi^-/\pi^+)_{132+124}/SR(\pi^-/\pi^+)_{108+112} }$, should primarily reflect the isovector mean fields that determine the symmetry energy. Experimentally, the double ratio cancels out most of the systematic errors but the statistical errors propagate. The current uncertainties in the double ratios shown in Fig.~\ref{fig:drspectra} are large and thus offer less precise constraints than single ratios. Nonetheless, the data are statistically consistent with the predictions indicated by the shaded area allowed by the 1-$\sigma$ range of the $L$ values (49-\SI{105}{MeV}) assuming the most probably value of $\Delta m^*_{np}/\delta =0.004$. 
 
  %The symmetry energy is expected to provide the dominant contribution to the pressure supporting the neutron star.  Based on the method outlined in~\cite{MORFOUACE2019135045, lynch2018nuclear}, we estimated that the pion spectral ratio is most sensitive to the symmetry energy at a density of $\approx 1.5 \rho_{0}$ even though maximum density of  $\approx 2 \rho_{0}$ can be reached during the collision. At that density, the present results suggest a representative symmetry pressure of $P_\text{sym}=  $$10.6\pm\SI{9.6} {\mega\electronvolt\per\femto\metre\cubed}$.
  
Additional measurements would reduce the uncertainties of this constraint. These include pion measurements at higher and lower incident energies to constrain non-resonant pion emission and the interactions of $\Delta$ baryons with nuclear matter. Precise measurements of the ratios of neutron and proton energy spectra should constrain $\Delta m^*_{np}$ more accurately 
  %than was achieved in Ref.~\cite{MORFOUACE2019135045}, 
  removing an important  contribution to the present uncertainty. Complementary measurements of proton and neutron elliptic flow are also desirable. Finally, ongoing efforts in transport theory by the Transport Model Evaluation Project (TMEP) collaboration (e.g. Ref.~\cite{Ono19}) would allow a more comprehensive exploration of the equation of state of dense neutron-rich matter via heavy ion collisions.   
  
In conclusion, we present precise spectra of charged pions produced in intermediate energy collisions involving rare isotope Sn beams on isotopic Sn targets and use them to constrain the symmetry energy at supra-saturation densities. To avoid complications resulting from poorly constrained $\Delta$ baryon potentials and non-resonant pion emission that are currently difficult to model, we focus our analyses on energetic pions with  $p_{T} >\SI{200}{MeV/c}$ and obtain symmetry energy constraints of $42 < L < \SI{117}{MeV}$ and $32.5 < S_0 < \SI{38.1}{MeV}$. These $L$ values are smaller than the values,  $L=106\pm\SI{37}{MeV}$ and $S_0=38.3\pm\SI{4.7}{MeV}$ ~\cite{reed2021implications}, extracted from a new measurement of the neutron skin thickness of $^{208}$Pb ~\cite{Adhik:21prex} though the two results are statistically consistent. Our results lie closer to the values $70 < L < \SI{101}{MeV}$ and $33.5 < S_0 < \SI{36.4}{MeV}$~\cite{Danielewicz_2017} extracted from charge exchange and elastic scattering reactions, and are larger than values  ~\cite{Yue21,essick2021astrophysical} influenced by NS deformability ~\cite{Abb17, Abb18} and radius ~\cite{Riley_2019,Miller_2019} measurements. 

\section{acknowledgement}

 The authors would like to thank Profs. Hermann Wolter, Che-Ming Ko and Pawel Danielewicz and the TMEP collaboration for many fruitful discussions via zoom during the pandemic. This work was supported by the U.S. Department of Energy, USA under Grant Nos. DE-SC0021235, DE-NA0003908, DE-FG02-93ER40773, DE-FG02-93ER40773, DE-SC0019209, DE-SC0015266, DE-AC02-05CH11231, U.S. National Science Foundation Grant No. PHY-1565546, the Robert A. Welch Foundation (A-1266and A-1358), the Japanese MEXT, Japan KAKENHI (Grant-in-Aid for Scientific Research on Innovative Areas) grant No. 24105004, JSPS KAKENHI Grants No. JP16H02179 and No. JP18H05404,
the National Research Foundation of Korea under grant Nos. 2016K1A3A7A09005578, 2018R1A5A1025563,2013M7A1A1075764, the Polish National Science Center(NCN) under contract Nos. UMO-2013/09/B/ST2/04064, UMO-2013/-10/M/ST2/00624, Computing resources were provided by the NSCL, the HOKUSAI-Great Wave system at RIKEN, and the Institute for Cyber-Enabled Research at Michigan State University. %A. Ono acknowledges support from Japan Society for the Promotion of Science KAKENHI Grants No. JP24105008 and No. JP17K05432.Y. J. Wang was supported by the National Natural Science Foundation of China under Grants No. 11847315, No. 11875125, and No. 11505057. H. H. Wolter was supported by the Universe Cluster of the German Research Foundation (DFG).
%

%\bibliography{references.bib}
\providecommand{\noopsort}[1]{}\providecommand{\singleletter}[1]{#1}%
%

% ****** End of file apstemplate.tex ******
\end{document}